\documentclass[12pt]{JHEP3}
\usepackage{amsmath}
\usepackage{amssymb}
\usepackage{amsthm}
\input xy
\xyoption{all}
\xyoption{rotate}

\newtheorem{Standard Theorem}[Theorem]{Standard Theorem}

\newtheorem{Counterexample/Proposition}[Theorem]{Proposition/Counterexample}

\newcommand\tr{\hbox{tr}}

\newcommand\BC{{\mathbb C}}
\newcommand\CAQ{{\cal A_Q}}
\newcommand\CO{{\cal O}}

\newcommand\BZ{{\mathbb Z}}

\author{Charlie Beil$^\dagger$, David Berenstein$^\ddagger$\\
$^\dagger$ Department of Mathematics, UCSB, Santa Barbara, CA 93106\\
$^\ddagger$ Department of Physics, UCSB, Santa Barbara, CA 93106\\
$^\ddagger$ Kavli Institute for Theoretical Physics China, CAS, Beijing 100190, China }

\title{Geometric aspects of dibaryon operators}

\abstract{The AdS/CFT correspondence for ${\cal N}=1$ Super conformal field theories suggests that dibaryon operators  are dual to D-brane states that are point like in AdS and that wrap various cycles in a Sasaki-Einstein manifold. It also suggests that the volume of the D-brane gives the R-charge of the corresponding operator. We elucidate various aspects of this correspondence, paying particular care to study the case of branes at the tip of three different Calabi Yau cones. We show that the arrows in the quiver diagram describing the conformal field theory can be thought of as global sections of a non-trivial holomorphic vector bundle over the Calabi-Yau geometry. We suggest that the zero locus of these sections gives the geometric map that lets us tie a particular dibaryon to a holomorphic cycle, by intersecting the corresponding cycle with the Sasaki-Einstein locus at fixed distance from the origin. We show that this can be compared with the corresponding volumes of the Sasaki-Einstein space and that one gets exact agreement between the volumes of the cycles identified with this procedure and the R-charges of the operators.}

\begin{document}

\section{Introduction and conclusion}
In recent years, it has been understood that the low energy physics of 
D-branes near Calabi-Yau singularities gives rise to supersymmetric field theories where the gravitational degrees of freedom can be decoupled. The field theory dynamics and the geometry of the Calabi-Yau singularity are very closely related. This relation is most striking in the AdS/CFT correspondence setup \cite{Malda}, where one ends up with a quantum field theory being exactly dual 
to a gravitational description of the near horizon geometry of the branes, meaning that they both describe the same dynamical system. The evidence for this identification is overwhelming and it suggests that for any object on the field theory side one can find a corresponding object in the gravitational side.

Given that the theories are supersymmetric, one can exploit holomorphy and the natural  geometry of branes to obtain identifications between field theories and geometries to provide long lists of conjectured dual pairs (for the case of toric singularities see \cite{FHMSVW}). If we have a local Calabi-Yau singularity $V$, we expect that it can be described by a three dimensional non-compact algebraic variety. One usually believes that this local problem will be characterized by some affine space: a commutative ring with polynomial relations, rather than a more formal algebraic geometric space that is built by patching these 
classes of objects into one bigger geometric object.  A similar holomorphic polynomial object is the chiral ring of a supersymmetric quantum field theory, so the two holomorphic structures should be related. One could ask how the commutative ring describing the variety $V$ is encoded in the supersymmetric quantum field theory. If moreover the field theory is conformal, the two rings will be graded and $V$ will be a complex cone.

The conformal field theory is usually considered to be a theory of $N$ point-like D-branes on $V$ and we usually expect that the moduli space of the field theory is given by $N$ points of 
$V$. However, when branes reach the singularity they fractionate: the singularity in moduli space is a physical singularity of the moduli space in the sense that new massless degrees of freedom should be present. This fractionation implies that the gauge symmetry of a single brane at the origin is not $U(1)$ (what one would expect for a single brane), but instead it is a product of 
gauge fields $\prod_i U(N_i)$, where the $N_i$ indicates the number of fractional branes of type $i$ that are present. One of the problems of establishing the $AdS/CFT$ duality is to determine these $N_i$ and the possible structures of the field theory given a singularity. In the toric case for a brane in the bulk, all the $N_i=1$, so the problem becomes simpler. However, for more general cases there is no uniform answer. 

If we blow-up the singularity, the fractional branes are expected to be some branes wrapping cycles in the exceptional locus with some bundles on them and one would call this construction a geometric engineering of a field theory \cite{Vafa}. There are also situations where one is not allowed to blow-up the singularity because one has discrete torsion degrees of freedom that prevent it \cite{VW}. This means that there can be more than one class of quantum field theories that gives rise to the same algebraic variety $V$. It would be nice if we could find a way to read $V$ directly from an 
abstract field theory even when we don't know how many branes of each type are needed to build a brane in the bulk (the $N_i$ are unknown). Such a program should also predict that the general form of the moduli space is some symmetric product of $N$ copies of $V$ (the branes are indistinguishable). This program were one begins with a field theory and produces the Calabi-Yau geometry where the branes live is the opposite process of geometric engineering, so one can label it reverse geometric engineering. An understanding of how $V$ appears in a given field theory has been described in \cite{Brev} (for previous work see \cite{BJL}), but one is not guaranteed to get a Calabi-Yau geometry from such a field theory and all the details of when this works are not understood in general. 

The basic idea in \cite{Brev} is that the field theories that appear are of quiver form and that one can associate to such a quiver theory with superpotential a non-commutative  associative algebra (ring) over the complex numbers. The ring is made of chiral gauge variant observables where one interprets the chiral matter fields of the 
quantum field theory as matrices that can be multiplied and added to make composite fields. One can then obtain gauge invariant observables by taking traces of these matrices. The advantage of this formulation in terms of algebra is that it produces a natural candidate for the variety $V$. The affine ring describing $V$ should be the set of elements of the algebra that belong to the center (in ordinary Lie algebra language these would be called the Cassimir-operators). This is a natural commutative subalgebra of any algebra. One also finds that a vacuum of the field theory is a representation of the algebra (independently of the rank of the gauge groups), so one can use this information to build general representations from the simplest ones and give a simple construction of the moduli space of vacua for arbitrary rank gauge groups. 

The traces of elements of the algebra being gauge invariant give natural elements of the chiral ring of the given field theory. The usual notion of the moduli space is that it is a representation of the chiral ring for a collection of fixed gauge groups \cite{Seiberg}, which can be quantum corrected by non-perturbative effects. The non-commutative algebra point of view is that the classical moduli space is a complete family of representations, which is given by direct sums of irreducibles ( a generalized notion of symmetric product), where the rank of the gauge groups is fixed once we fix a particular representative from the family.

One should  notice that when one goes to the conformal limit, the $U(1)$ parts of the gauge field decouple, either because they become massive
by mixing with closed string fields (this can happen if one cancels an anomaly via the Green-Schwarz mechanism), or because their coupling constant goes to zero. So in the end the gauge group is reduced
to $\prod_iSU(N_i)$ and the non-anomalous gauge $U(1)$ fields at zero coupling are left over as global symmetries. These act as generalized baryonic symmetries. This slight change of gauge group has implications for the field theory dynamics, in that we can expect that this decoupling of extra degrees of freedom might 
modify the moduli space of vacua and make it bigger, because one gauged symmetry becomes just a global symmetry, so the coordinates on this bigger symmetry space become physical.

This enlargement of the moduli space does not destroy the symmetric product structure: all the previous gauge invariant variables and the relations between them are the same. Thus the true moduli space must have a projection map to $Sym^N(V)$ and it gives it an extra lift. The coordinates describing the bigger moduli space should carry the extra $U(1)$ charges. These can be described by elements of the chiral ring that carry baryon charges. In general the possible baryons are not given by traces, but rather by determinant operators.
 Thus in order to understand this technicality one needs to improve the description of the moduli space. The purely non-commutative algebraic point of view does not know about baryon charges, because those depend on knowing the $N_i$ ahead of time.

One can also show that under the AdS/CFT dictionary, 
the traces that can be used to generate the chiral ring should be mapped to 
particular graviton states in the dual AdS theory \cite{WGKP}. If the dual theory is described by $AdS_5\times X$, for a Freund-Rubin \cite{FR} ansatz where $V$ is a real cone over $X$ and $X$ is a Sasaki-Einstein manifold \cite{MP},  it is expected that the baryonic operators are mapped 
to D-branes wrapping some non-contractible cycles in $X$ \cite{Wittenbar} (see also \cite{GK}). 

Mikhailov \cite{Mikhailov} has shown that the condition for the D-brane to be BPS 
can be related to studying holomorphic submanifolds of $V$ (this was the generalization described by Beasley in \cite{Beasley}). Where the holomorphic submanifold intersects the
Sasaki-Einstein manifold, we obtain a supersymmetric cycle that describes the instantaneous position of the brane at some time $t$.
The basic conjecture is that dibaryon and multi-baryon operators are given by some cycle inside $X$, but there is no obvious rule to follow that tells us which particular dibaryon or multi-baryon operator goes to which particular cycle of $X$. 

In a certain sense, natural conjectures as to how one should describe these objects are available in the literature \cite{IW}, but they are not explicit enough to provide a complete solution of the problem in the most general case. Also, they need to be corrected because it turns out that not all the matter fields joining two nodes correspond to the same anomalous dimension \cite{BBC}
(this results from the a-maximization procedure to calculate anomalous dimensions of fields \cite{IW1}).

We seem to be in a situation where the matching between operators and geometry is not obvious. The evidence for the identification is done by matching operator dimensions with the energies of branes in the gravity theory (one measures their volumes) plus any other quantum numbers that one has at ones disposal. So in situations where there is a lot of symmetry, one can in principle make a match of sufficiently symmetric objects. However, brane configurations can have degeneracies. This is because they come in families, and so we would like to know how to count these degeneracies as well as how to match them with the field theory. Moreover, we would like to have a precise map between field theory objects and geometric configurations, rather than a patchwork of matching of dimensions and degeneracies.

The purpose of this paper is to explain how to tie the conjectures between branes and dibaryons to a more precise formulation in terms of the (algebraic) geometry of the Calabi-Yau singularity that one wishes to study. In this way one can have a precise description that ties these two classes of objects (dibaryon operators and cycles) in a canonical way, rather than an ad-hoc prescription for matching dimensions and volumes of cycles. Our idea is to give an additional interpretation to the arrows of a quiver diagram as particular global sections of a non trivial holomorphic vector bundle over $V$. In two of the examples we study we have that the arrows can be considered as holomorhic global sections of line bundles over $V$, but we also study one more general example where this is not the case. Our techniques to show this depend on interpreting the quiver theory as a noncommutative geometric space as described above.
Our end result is similar to the prescription \cite{BFZ} that was done for the toric case. 
However, our techniques in the end do not depend on having this type of constraint on the quiver theory and they can be applied more generally.

Under these conditions, the zero locus of a section of a line bundle is a geometric object that can be identified with the holomorphic objects that Mikhailov's description requires. Moreover, we will see that the corresponding line bundles are determined by the coordinate ring of $V$: the locus where the vev of an arrow vanishes is a submanifold in the geometry determined by the coordinates of $V$.  These submanifolds are (Cartier) divisors corresponding to line bundles on $V$. If $V$ is a holomorphic cone with base $B$ it admits a $\BC^*$ action of rescalings. If the subvariety determined by the divisor of $V$ is invariant under the rescalings, then it also is a holomorphic submanifold of $B$.

 We will show how this works in detail for three different Calabi-Yau geometries.  We first consider the Klebanov-Witten setup of the conifold \cite{KW}, following the previous work of the second author \cite{BHK}.  The simplicity of this setup will provide intuition in how we approach our second more involved example, the Calabi-Yau complex cone over the first del Pezzo surface, which is the vev moduli space of the $Y^{2,1}$ gauge theory.  In this example we will be able to show that once we match the cycles in the Sasaki-Einstein manifold via the prescription we conjecture, the $R$-charges determined by $a$-maximization will be precisely the volumes of the corresponding zero-vev loci.  These nontrivial computations provide strong evidence for our conjecture. Finally, we will study a Seiberg dual realization of the orbifold geometry $\BC^3/\BZ_3$. In this geometry we show how to extend the results to more involved multi-baryon operators. The main difference with the previous two examples is that the arrows have to be interpreted as sections of some vector bundle of higher rank.  If one assembles the various arrows with care, one finds that one can generate a vector bundle map that is locally described by  a $k\times k$ matrix. The baryonic operators
end up being related to a locus where the map is degenerate: the kernel and cokernel jump rank. This is captured by the matrix. If we take the determinant of the matrix,  one can also be think about this more general object as the zero locus of a global section of the determinant line bundle associated to the map. However, in this case, the individual arrows will not have to vanish at the degenerate locus.

Our conclusion is that the procedure for identifying the geometry of dibaryons, although somewhat involved computationally in the examples,  can be made canonical from the point of view of algebraic geometry. It is also clear that this works in non-toric geometries and that we have a general setup for performing these computations. It should be interesting to apply these techniques to other conformal field theories and see the match between the geometry of  cycles and the counting of generalized dibaryons.

\section{Dibaryon operators in the conifold theory}

The simplest example to begin with is the conifold geometry. The conformal field theory was described by Klebanov and Witten \cite{KW} and derived by Morrison and Plesser \cite{MP}, and it consists of two gauge groups $SU(N_1)\times SU(N_2)$ with $N_1=N_2$ and a set of four bifundamental chiral fields $A_1, A_2$ transforming in the $(N_1, \bar N_2)$ and 
$B_1, B_2$ transforming in the  $(\bar N_1, N_2)$. The superpotential of the theory is given by
\begin{equation}
W\sim \epsilon_{ij} \epsilon_{lm}\tr(A_i B_l A_j B_m)
\end{equation}
where we use the natural contraction of indices as matrices to describe the superpotential. 
The $R$-charge of the fields $A,B$ is one half at the conformal point (the dimension of the scalar fields in the multiplet is $3/4$ rather than one).

The superpotential is invariant under an $SU(2)\times SU(2)$ global symmetry where the $A$ transform as a $(1/2,0)$ and the $B$ transform as a $(0,1/2)$. The superpotential is also invariant classically under a non-anomalous $U(1)_B$ baryonic symmetry where $A,B$ have equal and opposite charges. This charge is called the baryon charge. Indeed, one can think of the theories as being given by a $U(N)\times U(N)$ symmetry where only the $SU(N)$ part has been gauged. The extra non-gauged $U(1)$ symmetry is the baryonic number (there is a diagonal $U(1)$ which is decoupled).  The fields $A,B$ have a non-zero anomalous dimension when the theory is conformal. The field theory at the conifold should be considered strongly coupled.

One can show that the F-term equations lead to vacuum configurations where the $A_iB_j$ product matrices commute with each other. Generically, the $AB$ matrices do not have degeneracies in the 
eigenvalues and they can be diagonalized simultaenously by a $GL(N,\BC)$ transformation. In the superfield formulation, the gauge symmetry can be understood as acting by conjugation by $GL(N,\BC)$
or $SL(N,\BC)$ transformations, so this should be understood as an allowed gauge transformation \cite{WBagger}. The gauge invariant variables are the eigenvalues of the $AB$ matrices.

If we use the notation $U=A_1B_1$, $V=A_2B_2$, $W=A_1B_2$, $Z=A_2B_1$, we find that the
eigenvalues $u_i, v_i, w_i, z_i$ satisfy the equations
\begin{equation}
u_iv_i= w_i z_i\label{eq:conifold}
\end{equation}
so that each joint eigenvalue represents a point in the conifold, which is described by the equation
$uv=wz$. Thus, we find that the moduli space can be thought of a describing $N$ points in the conifold, one for each eigenvalue (see \cite{Bcon} for more details).

The eigenvalues of the $AB$ products can be recovered by taking traces of  products $\tr(U^{n_1} V^{n_2} W^{n_3} Z^{n_4})$. The order inside the trace does not matter because the matrices commute with each other. These traces are invariant under joint permutations of the eigenvalues. This permutation symmetry is a residual gauge symmetry of the system.

These operators are elements of the chiral ring, which is defined as the set of chiral gauge invariant scalar operators modulo the F-term relations. It is the F-term relations that guarantee that the order of the matrices does not matter within the chiral ring.

However, we have missed one important detail. We did the diagonalization at the level of the products $AB$, but we did not work the details of $A,B$ individually to see if there is more information we are missing. Indeed, there is such additional information that is not immediately apparent from the F-term relations. The key to understanding this information is that one can be more precise as to how one solves the moduli space of vacua problem. This is where one can see that there is a difference between gauging by $U(N)$ and by $SU(N)$ transformations in the gauge theory, \typeout{or by the complexification of the gauge group}
{or their respective complexifications} $GL(N,\BC)$ and $SL(N,\BC)$.
The missing generators of the chiral ring are going to be baryonic objects that are charged under the additional $U(1)_B$ symmetry. The traces defined above are neutral with respect to the $U(1)_B$ symmetry. 

For example, one can consider the objects given by
\begin{equation}
\det(B_1)(0) \sim \frac 1 {N!}\epsilon_{i_1\dots i_N} \epsilon^{j_1\dots j_N} (B_1)^{i_1}_{j_1}\dots (B_1)^{i_N}_{j_N}(0)
\end{equation}
This operator is invariant under $SL(N,\BC)\times SL(N,\BC)$ gauge transformations (the $\epsilon$ tensors are invariant tensors) and it is a baryonic object with charge $N$ (if each of the $B$ carry charge one). If we replace various $B_1$ by $B_2$, we find that this collection of operators transform as an $(0,N/2)$ representation of $SU(2)\times SU(2)$ \cite{BHK}. The symmetry is obtained by noticing that the $B$ are bosonic, and that the operator is totally symmetric in exchanges of the different $B$ objects. There are $N+1$ such operators. They are classified by how many $B_1$ and $B_2$ they have.

There are similar objects with $A$ fields rather than the $B$ fields. We can also consider a product of two or more of these operators. One important question is to count how many independent such products one can have. This counting can serve as a test of the $AdS/CFT$
correspondence if we can count states in the gravity side (one can sometimes do this via an index theorem for a quantization of a compact space of configurations).  We should always ask this question in the interacting theory. Another important question is to understand how to describe these objects in the dual gravitational theory in terms of the AdS/CFT correspondence. These issues have been studied in detail for the case of the conifold in \cite{BHK}, but a complete answer was not found for the multibaryon operators, nor for the correct counting of excited states of a given dibaryon. 
The problem of counting all chiral ring operators has been systematically started  in the works 
\cite{FHH}. 

This same type of reasoning can be applied to other \typeout{Conformal} conformal field theories in four dimensions. \typeout{If one has non-anomalous baryonic charges, one would like to know how to map various baryonic operators to geometric questions in the dual AdS theory.} In general, one would like to know the $AdS$ geometry corresponding to a given baryonic operator.

In the following subsection we outline the mathematical formalism that makes our prescription apparent.\typeout{We want to do some developments here that will make the description of the system better for our purposes. These should be considered as a mathematical formalism that will help in clarifying the form of the final answer and will make the geometric aspects more apparent.
Basically, we are going to} We describe the algebraic tools developed in \cite{BJL,Brev} for the 
three examples we consider in this paper.

\subsection{\typeout{Quiver theories as algebras together with representations of algebras.} Quiver gauge theories as representations of algebras}

This section reviews the points of view developed in \cite{BJL, Brev} regarding quiver theories as an algebra with an attached family of representations of the algebra. This mathematical point of view rewrites the problem of finding the solution to the moduli space of vacua into a problem of calculating the irreducible representation theory of an algebra. One can then assemble this information together into a full description of the moduli space of vacua that is easy to understand from the point of view of branes in some geometry. This is a generalization of the problem described above for a particular theory, where instead of commuting matrices one introduces a more general set of relations that need to be solved. In this description various aspects of the moduli space of vacua of a given field theory become obvious. In particular, the fact that the field theory moduli space is some generalized notion of symmetric product. Also, other aspects of brane fractionation can be characterized in a simple manner.

The idea is to begin with a quiver theory with some superpotential. A quiver is a graph with oriented arrows. The nodes of the graph will represent gauge groups and the arrows will represent matter fields.

 The gauge groups will be either $U(N)$ or $SU(N)$. The case of $U(N)$ is simpler and it does not have baryonic operators. We will be interested in the problem for the gauge group $SU(N)$. This will lead to a generalized version of symmetric product that takes into account baryonic operators. Once we understand how this works, we will be able to attack the problem of how to relate these operators to geometry in the dual gravity theory. The new development in this section is on how to get this additional information encoded into the algebraic setup.
 
In the superfield formulation of supersymmetric theories, the full quiver gauge group is really the complexification of $U(n)$, which is $GL(n)$.\typeout{In the supersymmetric theory, we should always consider the complexified gauge group associated to these gauge symmetries.} The extended gauge invariance resulting from the complexification is reduced to the \typeout{ordinary one,} usual non-complexified gauge invariance in the Wess-Zumino gauge. At the level of calculating the moduli space of vacua, the extra constraint is realized by noticing that the 
D-terms of the gauge theory \typeout{need to} must vanish. These D-terms are interpreted as moment maps in the full theory, so one can describe the 
problem of finding the moduli space of vacua as a symplectic quotient construction, which becomes equivalent to a geometric Invariant theory quotient. This is a fundamental result that will be used throughout this paper implicitly \cite{LT}.

The arrows in the quiver, joining nodes $i,j$ are in bifundamental representations $(N_i, \bar N_j)$ of the gauge groups associated to the end-points of the arrows. The directionality of the arrow indicates for which gauge group the matter is in the fundamental representation, and for which gauge group it is in the antifundamental representation. 

The natural two index structure of these objects makes it possible to think of these chiral fields as matrices that connect two auxiliary vector spaces of dimensions $N_i$ and $N_j$ over $\BC$. Let us 
call these auxiliary spaces $V_i$ and $V_j$. Thus, we can write in a formal way that
\begin{equation}
\phi_{ij} \in Hom(V_j, V_i)
\end{equation}
so that each arrow gives us a morphism between vector spaces. {We take the convention where $\phi_{ij}$ acts to the left of $V_j$, so for $v \in V_j$ we write $\phi_{ij}(v) \in V_i$.}\typeout{The convention here is that $\phi_{ij}$ acts to the left of $V_j$ and produces an element of $V_i$.} The arrows should be drawn according to this convention (the direction in which the \typeout{map acts} {morphisms act}).

\typeout{When we have matrices, there is natural way to multiply them if the index structure is just right.  Matrix multiplication in this language is just composition of morphisms.} {In this context, the composition of morphisms is represented by matrix multiplication.}  Obviously, we cannot multiply an 
arrow ending at node $j$ with an arrow beginning \typeout{in} at node $i$ in any non-trivial way that makes sense from the point of view of matrix multiplication, and so their product is defined to be zero.\typeout{It is still convenient to define this multiplication as zero.} 
To do this in a slightly formal way, \typeout{we want to introduce a collection of identity elements for each node. 
Let us call them $e_i$.}{we introduce an identity element $e_i$ for each node $i$.  Each $e_i$ may be viewed as a generator of the $U(1)$ gauge transformations at its corresponding node $i$, collectively normalized so that $e_ie_j = \delta_{ij}e_i$.  These ``idempotents'' additionally satisfy the relation} \typeout{One can think of these as a particular generator of $U(1)$ gauge transformations for the node $i$. The $e_i$ will be required to be normalized as follows $e_i^2 = e_i$, and $e_i e_j =0$ unless $i=j$.  Moreover, we get the identity}
\begin{equation}
1 = \sum_i e_i
\end{equation}

The statement that arrows begin and end on given nodes can be formalized in terms of algebraic matrix equations of the following form
\begin{equation}
\phi_{ij} e_k = \delta_{jk} \phi_{ij}, \quad e_k\phi_{ij} = \delta_{ik} \phi_{ij} 
\end{equation}
while the 'gauge' transformations by the $U(1)_i$ would be given by commutators with $e_i$. It is easy
to convince oneself that these rules make sense.

Now, we can ask what is the role of $GL(N_i,\BC)$ and $SL(N_i,\BC)$ transformations from this more formal matrix point of view. Well, elements of $GL(N_i,\BC)$ or $SL(N_i,\BC)$ act on a natural way on 
the vector spaces $V_i$. They do specific changes in the basis of $V_i$. When we change the basis of $V_i, V_j$ we can think of the matrices $\phi_{ij}$ as being invariant objects that do not depend on a basis, but their specific components do transform with the change of basis. These will be identical to the gauge transformations of the fields $\phi_{ij}$ if we are careful. Thus, we realize that the fields $\phi_{ij}$ transform covariantly with respect to this auxiliary structure.

The other thing we notice is that composition of matrices is associative, so these multiplications of fields to make composite fields can be encoded naturally into a framework of having an associative product for the fields. The idea is that now we can abstract these concepts to state that the fields $\phi$ have a natural multiplication on their own, even in the absence of the vector spaces $V_i$ or the labels $N_i$.
This is, the fields $\phi$ in this situation give rise to a natural algebra structure on their own. The generators of the algebra are the arrows of the quiver and the $e_i$, and all composite paths of arrows define 
abstractly an element of the algebra. We will allow to take general finite linear combinations of these 
objects with complex coefficients. The end result we get is a path algebra of the associated graph.

We are just showing that the quiver structure associated to gauge theories of a particular type naturally lead to a notion of an associative algebra. Let us name this algebra ${\cal A_Q}$, where ${\cal Q}$ is the quiver. So what happens when we substitute arrows for specific matrices? In the algebraic setup, this means that we have a map $\mu:\phi_{ij}\to Hom(V_j,V_i)$ where we have prescribed some particular vector spaces $V_i$, and 
such that the abstract definition of the algebra of the $\phi$ with it's tautological (standard) multiplication is reflected into having matrices $M_{ij}$ that satisfy the same multiplication table. At this level, this is essentially trivial, because we have essentially no non-trivial relations between the generators, but from a formal point of view what we realize is that we have a representation of the algebra ${\cal A_Q}$ realized by matrices. This is, given the $V_i$, any collection of matrices will do, where the only ones that are fixed are $e_i$. They are such that
$e_i v_j = \delta_{ij} v_j$ for any vector $v_j \in V_j$.

So far, the algebra we have is very easy to understand. We have relations from incidence into the different nodes of the quiver and that is all. The idea is that now we can consider a superpotential for the field theory.

Within perturbative string theory one usually generates a superpotential of the general single trace form (a disc diagram on the worldsheet). These are the natural objects that can be associated to geometry.
The superpotential will have the general form
\begin{equation}
W= \tr( X )
\end{equation}
where $X$ is any element of the path algebra and $\tr$ stands for an ordinary matrix trace (this is invariant under cyclic permutations and also under similarity trasnformations). One can show easily that $X$ can only depend on {oriented cycles (closed paths) in the quiver}\typeout{closed paths}. This is because
\begin{equation}
W= \tr (1^2 X)=\tr( (\sum e_i)^2 X)= \sum\tr (e_i^2 X)= \sum \tr( e_i X e_i)
\end{equation}
Notice that $e_i Xe_i$ are paths that begin and end on node $i$. This condition tells us that the superpotential ends up being gauge invariant, because on each such \textbf{cycle}\typeout{closed path} the $SL(N_i,\BC)$ change of basis acts by conjugation and the trace is invariant due to the cyclic property. 

If one considers the F-term equations associated to this superpotential, it is clear that they can also be written as algebraic relations involving the generators in the associative algebra 
\cite{BJL, Brev}. Thus, what we have is a path algebra with relations and the relations are
derived from a superpotential.

For example, in the case of the conifold, the F-term relations read
\begin{eqnarray}
B_1 A_i B_2&=& B_2 A_i B_1\\
A_1B_i A_2&=& A_2 B_i A_1
\end{eqnarray}
This means that the problem of solving the F-term constraints reduces to the problem of finding matrices that satisfy these relations, modulo gauge transformations (which at the level we have described are a change of basis of the $V_i$). This identification under change  of basis can be thought of as equivalence classes of representations, so they are objects in the category of modules of the algebra $\CAQ$. It is easy to manipulate the above equations to show that
\begin{eqnarray}
W_{ij} W_{lm} = W_{lm} W_{ij}
\end{eqnarray}
where $W_{ij} = A_i B_j$. Similarly, we can consider the $\tilde W_{ij} = B_j A_i$ and we can show that these also commute with each other.

One can also use the more formal objects of the algebra $Z_{ij} = W_{ij}+ \tilde W_{ij}$. It is easy to show that 
$Z_{ij}$ commutes with all of the elements of the algebra. We do this by showing that it commutes with the $e_i$ and $A_i, B_i$, the generators of the algebra.  This is a simple exercise \cite{Bcon}.

One can also recover the original $W, \tilde W$ matrices by projecting with the $e_i$ as follows  $W_{ij} = e_1 Z_{ij} e_1$ and $\tilde W_{ij} = e_2 Z_{ij} e_2$. 

Finally, one also gets the relations $Z_{11} Z_{22} = Z_{12} Z_{21}$ which is a matrix version of the conifold geometry (\ref{eq:conifold}).

Now, let us assume that we have two representations of the algebra and let us call them $R_1$, $R_2$. It is easy to show that $R_1\oplus R_2$ is also a representation. This is the standard direct sum of modules for the algebra. Another standard result of representation theory is that if one has a module map between representations of an associative algebra
\begin{equation}
\mu: R_1\to R_2 
\end{equation}
then the kernel and the coset $R_2/\mu(R_1)$ are also representations. This means that we can build more general solutions of the relations by using smaller representations. 

A representation $R$ is called irreducible if it has no subrepresentation inside it. 
This is, if one has any map 
$\mu: R_1\to R$, and $\mu(R_1)\neq 0$, then $\mu(R_1)=R$.

For irreducible representations one can use Schur's lemma. This states that any element of the center is proportional to the identity. We have already seen that the $Z_{ij}$ all belong to the center. This means that in an irreducible representation they should be proportional to the identity. Notice that the $Z_{ij}$ define a commutative algebra over the complex numbers, and that they recover the conifold 
geometry as the algebraic variety defined by the center algebra.

Also, since the $e_i$ form a complete set of projectors that commute with each other, we can choose representations where the $e_i$ are diagonal. These are of the form
\begin{equation}
e_1 \sim \begin{pmatrix}1&0\\0&0\end{pmatrix}, e_2\sim \begin{pmatrix}0&0\\0&1\end{pmatrix}\label{eq:gen}
\end{equation}
where the block diagonal decomposition can be of arbitrary dimension.

If we also write the algebra carefully, we find that as a left module over the center, the algebra is generated by $A_i, B_i, e_1, e_2$. This is, a general algebra is finite dimensional over its center.  Thus the full algebra of the quiver theory can be interpreted as a particular coherent sheaf of finite dimension over the usual algebraic variety (this turns out to be a holomorphic vector bundle away from the singularity).

Working a little bit harder, one finds that the irreducibles can be described with $2\times 2$ matrices where the $e_i$ are given as in (\ref{eq:gen}), and the $A,B$ matrices are strictly off-diagonal. There are no relations between them. So, a naive guess is that the moduli space is $\BC^4$. However, if we gauge the $GL(1,\BC)$ symmetry, we get that the moduli space is
properly $\BC^4/\BC^*$ and the conifold is a quotient variety. The extra dimension 
we get in the moduli space should be thought of as a baryonic direction (this has been called the Master space \cite{Master}).  Also, given any representation, the dimensions of $V_1, V_2$ can be calculated by taking traces of $e_1, e_2$ and in this case this gives us one for each.

Under these conditions, the $A$ become just numbers, as well as the $B$: they act as homomorphisms between one dimensional vector spaces.

The exception to this happens only if none of the $Z_{ij}$ are invertible, where one finds smaller representations, with $e_1=1, e_2=0$, or $e_2=1, e_1=0$ and all arrows given by zero. Let us call these $S_1, S_2$. These smaller representations at the singular locus can be called fractional brane representations. One can show that if we take the general representation $R$ of $2\times 2$ matrices and we let $A\to 0, B\to 0$, then we have
\begin{equation}
\lim_{A,B\to 0} R \to S_1\oplus S_2
\end{equation}
while one has a non-trivial short exact sequence
\begin{equation}
0\to S_1\to \lim_{A\to 0} R \to S_2\to 0
\end{equation}
that is parametrized by the values $b_2,b_1$. This makes the identification of arrows in the quiver with 
$Ext$ groups easy to understand. The $Ext^1(S_2,S_1)$ groups can be characterized by the equivalence classes of such extensions (this is the counting of massless modes between D-branes \cite{Douglas}). In this case, the dimension space of the extension space is two. This is the same as the number of arrows in the quiver diagram going from node one to node two.
One can do a similar analysis with the other arrows, by exchanging the roles of $S_1, S_2$.

So far we have been cavalier with the role played by $GL(N,\BC)$ or  $SL(N,\BC)$. At this level we have not encountered an obvious difference yet. The difference is at the level of which changes of basis are allowed. If we allow general changes of basis, then we are working with $GL(N,\BC)$. However, to work with $SL(N,\BC)$ we end up with a restriction on the changes of basis: we are only allowed to make a change of basis that preserves a volume form for the  $V_i$. This is, we have to choose a preferred element of $\omega_1\in \Lambda^N V_1\simeq \BC$ and $\omega_2\in\Lambda^N V_2\simeq \BC$, which are one dimensional vector spaces over $\BC$. This is not something that fits easily within a purely algebraic problem. 

However, we can work around this by using a compensator. We can allow general changes of variables, so long as we compensate by rescalings of the volumes as we change variables. This means that for theories with $SL(N,\BC)$ groups we can still use the general changes of variables, but we have to add volume forms on the vertices. These volume forms also transform. If we ignore the volume forms, we get back the same problem as with 
$GL(N,\BC)$ group, and that has been solved in terms of representation theory 
already. 

If we put many 
of these representations together, and we gauge the $SL(N,\BC)$ symmetry, we will get that
the total moduli space is a line bundle over $N$ copies of the conifold. This is because the action of the $SL(N,\BC)$ gauges the $GL(1,\BC)^{N-1}$ diagonal subgroup. In this manner we find that
\begin{equation}
{\cal M}_{GL} = {\cal M}_{SL} //GL(1,\BC)
\end{equation}

The fact that the representation theory still works means that the full moduli space in the case of the $SL(N,\BC)$ theory can be mapped onto the moduli space of the $GL(N,\BC)$ theory in a canonical way, and we have an algebraic fibration structure 
\begin{equation}
GL(1,\BC)\to {\cal M}_{SL} \to {\cal M}_{GL} 
\end{equation}

The fibration is parametrized by the (complex) size of the volume forms, so it should be thought of as a $(\BC^*)^2$ fibration, this is the same as having $GL(1,\BC)^2$ orbits.  Now, a map like $A_1$, from $V_1$ to $V_2$ also acts on the volume forms in an obvious way (the matrix $A_1$ can act by products on tensors). We can call the quantity
\begin{equation}
\det(A_1) \sim A_1( \omega_1)/ \omega_2
\end{equation}
and this gives our notion of dibaryon operators. This is a number on a given representation with
choices of volume forms. These are the new coordinates of the chiral ring.

Since the moduli space for the $SL(N,\BC)$ theory is naturally fibered over the moduli space for the 
$GL(N,\BC)$ theory and the fiber is finite dimensional, while the base can have arbitrarily large dimension, it makes sense to try to think geometrically in terms of the natural geometry of the base.
The base is the moduli space of vacua of the $U(N)\times U(N)$ theory. It is a symmetric product  $Sym^N(V)$, where $V$ is the conifold variety, the locus $uv=zw$ in $\BC^4$.
Each of the points of $V$ that is selected is described by an irreducible representation. 

What we would like to have now is a geometric interpretation of the fields $A, B$ in terms of the geometry of the conifold, even if it is on an irreducible representation. So far, this is not apparent. This is where having the algebraic description given above will make a difference.

Consider the algebra of the conifold quiver ${\cal A_Q}$. The algebra can be considered as
a left module over itself: we multiply by elements of the algebra on the left.
The algebra can be split as a left module in the obvious form
\begin{equation}
{\CAQ} = \CAQ e_1 \oplus \CAQ e_2
\end{equation}
This is because every arrow in the quiver ends in one of the two nodes. Indeed, if we multiply 
by elements of $\CAQ$ on the right, this commutes with left multiplication, so it provides a 
natural way to build module maps. If we multiply by $1= e_1+e_2$ on the right we recover the splitting above.
Each of the modules $\CAQ e_1$ and $\CAQ e_2$ are direct summands of a free module ($\CAQ$ itself) and therefore they are projective. Projective modules are the natural generalization of vector bundles in this context. 

Each of these is also finitely generated over the center of the algebra (the algebra itself has that property). Thus, one should be able to think of $\CAQ e_1$ and $\CAQ e_2$ as vector bundles over the conifold geometry. Since the conifold geometry is singular, one can localize this property away from the singular locus $u=v=w=z=0$ by taking inverses for some of these 
variables.

For example, we can consider the complex submanifold described by $u$ being invertible.
We can consider a different patch where $w$ is invertible. 
If $u$ is invertible, the quotient $w/u$ makes sense. Similarly, if $w$ is invertible, the quotient $u/w$ makes sense. With these two patches one can construct the blow-up of the conifold at the origin. The coordinates $\xi= w/u$ and $\xi'=u/w$ are patched as $\xi'= 1/\xi$. These
describe a $\mathbb{CP}^1$ geometry. The blow up of the conifold is the total space of $\CO(-1)\oplus\CO(-1)$ over $\mathbb{CP}^1$. The coordinates $\xi, \xi'$ describe the coordinates of the base of this fibration.

Notice that on an irreducible representation we also have that
\begin{equation}
u/w = Z_{11}/Z_{12} = b_1/b_2
\end{equation}
but in the full algebra one should write instead
\begin{equation}
Z_{11}/Z_{22} = e_1 B_1  B_2^{-1} e_1 + e_2 B_2^{-1} B_1 e_2
\end{equation}
because the $B$ are not invertible in the full algebra, but they are as maps between the 
$V_i$.

For an irreducible representation the quotient of the arrows can be thought of as a set of coordinates on $\mathbb{CP}^1$. The non-normalized coordinates $b_1, b_2$ can be interpreted as two distinct holomorphic sections of the hyperplane bundle on this $\mathbb{CP}^1$. Notice also that the locus where $b_1$ vanishes does not depend on the normalization of the arrows and it defines a holomorphic submanifold of the blow-up. We can project this submanifold to the blowdown and we can therefore identify some 
geometric locus on the conifold $V$ associated to an arrow of the quiver itself. This is the locus where $b_1$ vanishes. We can do the same for the $A$ arrows.

In essence, from the point of view of geometry, we can interpret the noncommutative variables that extend the commutative geometry of the center to include non-diagonal matrices as
holomorphic sections of particular (line) bundles on the complement of the singular locus. 
These also belong to 
\begin{equation}
Hom(\CAQ e_1, \CAQ e_2)
\end{equation}
 because of our interpretation of arrows in the quiver as maps between these modules. These 
 $Hom$ functors are also modules over the center and define for us a coherent sheaf on the conifold, whose global sections are the paths in the quiver starting on one node and ending on another one. A particular choice of an off-diagonal element corresponds to a particular choice of global section.

For the case of toric quiver diagrams the elements of $Hom(\CAQ e_1, \CAQ e_2)$ describe a line bundle. This is because one expects that the set $e_i \CAQ e_i$ is isomorphic to the center of the 
algebra. Thus, if we take the two non-zero elements $v_1, v_2\in Hom(\CAQ e_1, \CAQ e_2)$, we can compare them by taking quotients (the ratio $v_1/v_2$ makes sense at a generic point) and gives a rational function of elements of  $e_1 \CAQ e_1$: the algebraic variety describing the singularity. This is the same type of comparison that tells us that we have a line bundle: we produce rational functions by taking ratios of global sections. The divisor characterizing the line bundle is the polar locus of the quotient. 
 
 We can also blow up the $A$ variables. This gives us a different $\mathbb{CP}^1$. Between the $A$ and the $B$ variables we can see that there is a ${\mathbb{CP}^1}\times {\mathbb{CP}^1}$ space appearing. The Sasaki-Einstein space $T^{11}$ is  a regular circle fibration over $\mathbb{CP}^1\times \mathbb{CP}^1$.
 This $\mathbb{CP}^1\times \mathbb{CP}^1$ is the complex base of the Calabi-Yau cone.
 
 The dibaryon operators in $AdS_5\times T^{11}$ are branes that are located at one point of either of the $\mathbb{CP}^1$. They wrap the other $\mathbb{CP}^1$ and the circle fiber. This is the intersection of the locus $a_1=0$ or $b_1=0$ with the Sasaki-Einstein base. 
 The locus $a_1=0$ is the same locus as $\{(u,v,w,z) \in \mathbb{C}^4 \ | \ uv-wz=0 \text{ and } u=0 \cap w=0\}$. Remember that $u=a_1b_1$, $w=a_1b_2$, so that $a_1$ vanishing implies that both $u,w$ have to vanish.
 
 Notice that in the conifold, even though this is a space of codimension one, it can not be described as the zero locus of a single holomorphic function (like setting $u=0$ alone). This is how we know algebraically that one needs a section of a nontrivial line bundle over the conifold in order to describe it.

\subsection{Conjectures}

To summarize, our algebraic point of view has upgraded the chiral fields in a quiver with a superpotential to be related to global sections of some holomorphic vector bundles (coherent sheafs) on the geometry of the center algebra, which  we will take to be the Calabi-yau geometry where D-branes are moving (see \cite{Brev} for conjectures regarding how these ideas fit together into one framework and what conditions are required for this point of view 
to be useful).

{The zero locus of a given global section defines a holomorphic submanifold of the Calabi-Yau geometry. Mikhailov has shown that in conformal field theories, holomorphic submanifolds can be used to define dibaryons and giant gravitons: they are the locus where the holomorphic submanifold intersects the Sasaki-Einstein base of the Calabi-Yau geometry, which is at a fixed distance from the origin. The Sasaki-Einstein manifold is itself a circle bundle whose base
 is a projective variety.}
 
\typeout{The zero locus of a given global section defines a holomorphic submanifold of the Calabi-Yau geometry. Now, for conformal field theories, Mikhailov has shown that on general grounds holomorphic submanifolds 
 can be used to define dibaryons and giant gravitons: they are the locus where the holomorphic submanifold intersects the Sasaki-Einstein base of the Calabi-Yau geometry, which is at fixed distance from the origin. A Sasaki-Einstein manifold is a complex cone whose base
 is a projective space.}
 
 For the case of the conifold, the base is a $\mathbb{CP}^1\times \mathbb{CP}^1$ geometry, and the Sasaki-Einstein manifold is a circle bundle over the $\mathbb{CP}^1\times \mathbb{CP}^1$ base. If we take a homogeneous global section (with fixed R-charge), the zero locus will be invariant under the 
 $U(1)_R$ symmetry of the conformal field theory and it will wrap the circle direction of the Sasaki-Einstein space. This locus can be projected down to the $\mathbb{CP}^1\times \mathbb{CP}^1$ base and we get a holomorphic object on the base of the cone.

 The idea is that we will associate exactly such a geometric locus to any 
 element of the algebra that begins in one node and ends at another (possibly the same) 
 node. This conjecture can be made more convincing with the ideas of emergent geometry as espoused in \cite{BHart}. One understands that the chiral ring is a holomorphic quantization of the 
moduli space of vacua. A holomorphic quantization would require a line bundle ${\cal L}$ on the moduli space of vacua, and the wave functions would be holomorphic sections of such a line bundle. If we just look at the symmetric product base, one usually takes the structure sheaf bundle for ${\cal L}$, but this is not required. Since the moduli space is a fibration over the symmetric product base, a dibaryon wave function is sensitive to the details of this fibration, and this would result in holomorphic sections over the base that are due to different line bundles. These would measure the twisting of the fibre by the dibaryon charge. This means that each dibaryon (or multi-baryon) charge  picks a line bundle and a specific dibaryon-like operator would pick a global section of the associated line bundle.  A similar prescription can be found in \cite{BFZ} which specializes to the case of toric geometries and the line bundles are constructed from the toric data. 

As we have seen, for the case of the conifold, this a straightforward procedure and it matches the previous known geometric results exactly. Now we want to apply this to the case of the 
Calabi-Yau manifold called $Y^{2,1}$. This is a non regular Sasaki-Einstein manifold and its metric was constructed in \cite{GMSW}.

\section{An example: the $Y^{2,1}$ quiver gauge theory}

 The $Y^{2,1}$ quiver gauge theory has quiver
$$
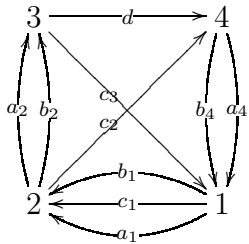
\xy (0,0)*+{2}="1";(0,25)*+{3}="2";(25,25)*+{4}="3";(25,0)*+{1}="4";
{\ar@/^/|-{a_2}"1";"2"};{\ar@/_/|-{b_2}"1";"2"};{\ar@/^/|-{a_4}"3";"4"};{\ar@/_/|-{b_4}"3";"4"};
{\ar@{->}|-{d}"2";"3"};{\ar@/^1pc/|-{a_1}"4";"1"};{\ar@/_1pc/|-{b_1}"4";"1"};
{\ar@{->}^{c_3}"1";"3"};{\ar@{->}_{c_2}"2";"4"}; {\ar@{->}|-{c_1}"4";"1"};
\endxy$$
and superpotential
\begin{equation} 
\label{hi} W = \text{tr}\left(c_3 \left( b_2a_1 - a_2b_1 \right) + c_1 \left( b_4da_2 - a_4b_2 \right) + c_2 \left( b_1a_4 - a_1b_4 \right)\right).
\end{equation}
The theory was constructed in \cite{FHHtoric} (see also \cite{HW} for a different viewpoint), and has been analyzed extensively in many papers. We will follow some of the algebraic geometric details described in \cite{BHOP} to make everything as explicit as possible.

It has been proven in \cite{B1} that a single brane at a generic point of the vacuum moduli space (that is, an irreducible representation of the corresponding algebra) will have a 1 dimensional vector space at each node and so the arrows will be represented by complex numbers; this is expected since the $Y^{2,1}$ geometry is toric.  The gauge group will then be $\prod_{1 \leq i \leq 4}GL(1,\BC)$, and consequently only the traces of fields (representations of cycles) count as gauge invariant observables. These can be characterized by the set of cycles centered at any of the nodes. It is a nontrivial task to show that if one solves the F-term constraints, then the set of cycles centered at one node are related to the set of cycles centered at any other node, and their representations give identical numerical values (this has also been proven in \cite{B1}). 

As we will show below, the vacuum moduli space arising from the quiver (the moduli space of irreducible representations of the algebra) is a complex cone whose base is the first del Pezzo surface $dP_1$, which is $\mathbb{CP}^2$ blown-up at one point.  It is conjectured that the moduli space is also a real cone over the $Y^{2,1}$ non-spherical horizon.  

We will first describe the algebraic geometry of the $dP_1$ surface in detail. Then we will see how we can read the $dP_1$ geometry from studying the quiver algebra. The idea is that
$dP_1$ can be embedded in a simple way into the algebraic variety 
$\mathbb{CP}^2\times \mathbb{CP}^1$.
Using the projective coordinates $\left[x_1: y_1: z_1; x_2:y_2 \right]$ for the $\mathbb{CP}^2\times \mathbb{CP}^1$ space \footnote{In this notation, $x_1, y_1,z_1$ cannot all be zero and $x_2, y_2$ cannot both be zero.  Also for any $\lambda \in \mathbb{C}^*$,
$$\left[ \lambda x_1: \lambda y_1: \lambda z_1; x_2:y_2 \right] \sim \left[ x_1:y_1:z_1; x_2:y_2 \right],$$
$$\left[ x_1: y_1:z_1; \lambda x_2: \lambda y_2 \right] \sim \left[ x_1:y_1:z_1; x_2:y_2 \right].$$}
$dP_1$ may be described as the locus $\{ x_2y_1 = y_2x_1 \} \subset \mathbb{CP}^2 \times \mathbb{CP}^1$.
The $dP^1$ has a clear projection to $\mathbb{CP}^2$ which is done by using the forgetful map
that sends $f: [x_1:y_1:z_1; x_2:y_2] \to [x_1:y_1;z_1]$. This map is one to one away from $f^{-1}[0,0,1]$. The exceptional divisor of $dP^1$ is the inverse image of $[0,0,1]$, and
we call it $E= f^{-1}[0,0,1]$. This inverse image is a copy of $\mathbb{CP}^1$ (the equation
that the $dP^1$ variety satisfies is solved for arbitrary values of $[x_2:y_2]$).

If we consider an hyperplane $H$ on the $\mathbb{CP}^2$ projection that does not pass through $[0,0,1]$, then the inverse image $f^{-1}(H)$ will be a subvariety of $dP^1$ which we will also call the hyperplane locus. It is easy to establish the following intersection numbers $H\cdot E=0, H\cdot H=1, E\cdot E=-1$. 
The subvarieties $E,H$ generate the second homology group of $dP_1$. It is also convenient to introduce a cycle which we will call a line $L$. The immersion on $\mathbb{CP}^2\times \mathbb{CP}^1$ has a projection $\pi$ onto the $\mathbb{CP}^1$ (again defined by a forgetful map onto the second set of coordinates). If we take any point $p \in \mathbb{CP}^1$, we will define the class of $L$ as $\pi^{-1}(p)$.

It is easy to establish that $L$ projected onto $\mathbb{CP}^2$ gives a line passing through the point that is the blow-down of the exceptional divisor (we will call this point the origin in $\mathbb{CP}^2$). One easily shows then that $L\cdot E=1$, $L\cdot H=1$, $L\cdot L=0$. But the class of $L$ should be a class in the second homology of $dP^1$, which is generated by $H,E$. One can determine that $[L]=[H]- [E]$. Thus, it will be better for us to think of $H= L+E$ in homology. Also, if one has a Kahler metric on $dP_1$, we will find that $Vol(H)= Vol(L)+Vol(E)$. 

Now, let us consider the complete linear system associated to the divisor $3H-E$ (this is a line bundle such that the zero locus of any of it's global sections is in the class $3H-E$ ). The complete linear system is a linear space made of all the holomorphic sections of a given line bundle. If the dimension of this linear system is $d$ (the number of linearly independent holomorphic sections), this complete linear system can be used to define a Veronese map from $dP^1$ to $\mathbb{CP}^{d-1}$, where the embedding coordinates are the values of the global sections at each 
point and they are mapped to the homogeneous coordinates of $\mathbb{CP}^{d-1}$. The multiplicative changes due to how one patches a line bundle together cancel because
the homogeneous coordinates of $\mathbb{CP}^{d-1}$ are well defined only up to multiplications by a common factor.

If we project the cycle $3H-E$ onto $\mathbb{CP}^2$, the fact that $[3H-E]\cdot E=1$ implies that the projection must pass through the origin. Moreover $[3H-E]\cdot H= 3$, so the projection gives a degree $3$ curve in $\mathbb{CP}^2$. This means that the curve is characterized by the zero locus of a cubic equation in $x_1, y_1, z_1$. Let us call 
\begin{equation}
S= \sum_{i+j+k=3, i,j,k\geq 0} a_{ijk} x_1^iy_1^jz_1^k
\end{equation} 
the associated equation.
Since $S$ passes through the origin, it is such that $S[0,0,1]=0$, and this tells us that $a_{003}=0$.
 
Any such curve can be lifted to $dP_1$ uniquely. Since $dP_1$ and $\mathbb{CP}^2 $ are birationally equivalent, if we take $\mathbb{CP}^2$ and remove the origin, we can describe the Veronese  embedding due to this linear system on $\mathbb{CP}^2 \setminus \{0\}$. It's completion in $dP_1$ will be the associated map. The global sections are the monomials of $S$ that are allowed.
This gives us a system of dimension $8$. This means that there is a natural embedding of $dP^1$ into 
$\mathbb{CP}^8$ and that the embedding is of degree $3$ for the hyperplane bundle. This is a standard construction in algebraic geometry \cite{GH} and it will make an appearance in the $Y^{2,1}$ field theory. The embedding is one to one.

According to the conjectures we have described, the locus where  a bifundamental field vanishes is a submanifold of the base.  The $R$-charge of the field is then the volume of this submanifold with respect to a Kahler-Einstein metric \cite{Wittenbar,GK} (see also \cite{BHK,IW}).  To test the conjecture for a given field we first find the zero locus of its vev and show that the volume of this locus is exactly the fields $R$-charge as computed by $a$-maximization.

For concreteness we will consider the minimal cycles at node 1 (again, the results are independent of what node is chosen), where a minimal cycle is one which has no proper cyclic subpaths.  Using the constraints 
 $\partial_a W = 0$, it was shown in \cite{BHOP} that every minimal cycle at node 1 is equal to one of the following minimal cycles:

\begin{equation}\begin{array}{ccc}
\varphi_{00} &= & a_4c_2a_1\\
\varphi_{01} &= & b_4c_2a_1\\
\varphi_{02} &= & b_4c_2b_1
\end{array} \ \ \ 
\begin{array}{ccc}
\varphi_{10} &= & a_4da_2a_1\\
\varphi_{11} &= & b_4da_2a_1\\
\varphi_{12} &= & b_4db_2a_1\\
\varphi_{13} & =& b_4db_2b_1
\end{array} \ \ \ 
\begin{array}{ccc}
\varphi_{20} & =&c_3a_2c_1\\
\varphi_{21} &= &c_3b_2c_1
\end{array}
\end{equation}
 Given a generic point in the vev moduli space (irreducible representation), it has been shown in \cite{B1} that there exists a choice of gauge (isoclass representative) for which
\begin{equation}
\left\langle a_2 \right\rangle = \left\langle a_4 \right\rangle, \ \ \ \left\langle b_2 \right\rangle = \left\langle b_4 \right\rangle, \ \ \
\left\langle c_1 \right\rangle = \left\langle c_2 \right\rangle = \left\langle c_3 \right\rangle, \ \ \ \left\langle d \right\rangle = \frac{\left\langle a_1 \right\rangle}{\left\langle a_2 \right\rangle} = \frac{\left\langle b_1 \right\rangle}{\left\langle b_2 \right\rangle}.
\end{equation}
Let us call the variables as follows
\begin{equation}\left\langle a_{2i+1} \right\rangle =: x_1, \ \ \ \left\langle b_{2i+1} \right\rangle =: y_1, \ \ \ \left\langle c_i \right\rangle =: z,\end{equation}

\begin{equation}
\left\langle a_{2i} \right\rangle =: x_2, \ \ \ \left\langle b_{2i} \right\rangle =: y_2, \ \ \ \left\langle d \right\rangle = \frac{x_1}{x_2}=\frac{y_1}{y_2}.
\end{equation}

Using this notation and setting $\phi_{jk}:= \left\langle \varphi_{jk} \right\rangle$, the vev's of the above minimal cycles are as follows:
\begin{equation}\begin{array}{ccc}
\phi_{00} &= & x_2 z x_1\\
\phi_{01} &= & y_2zx_1\\
\phi_{02} &= & y_2zy_1
\end{array} \ \ \ 
\begin{array}{ccc}
\phi_{10} &= & x_2^2x_1\\
\phi_{11} &= & y_2x_2x_1\\
\phi_{12} &= & y_2^2x_1\\
\phi_{13} & =& y_2^2y_1
\end{array} \ \ \ 
\begin{array}{ccc}
\phi_{20} & =&x_2z^2\\
\phi_{21} &= &y_2z^2
\end{array}
\end{equation}

At the apex of the cone the vev of each $\phi_{jk}$ is zero, and using the $F$-constraints we find that if we are away from the apex, not all $x_1,y_1,z$ can be zero and not both $x_2,y_2$ can be zero.  The base of the cone ``vanishes'' at the apex, so if we want to consider the base we must require not all $x_1,y_1,z$ be zero and not both $x_2,y_2$ be zero.  
There is one $F$-constraint between these coordinates, namely $0=\partial_{c_3}W = a_2b_1-b_2a_1$.  The resulting locus is then $\{x_2y_1=y_2x_1 \} \subset \mathbb{CP}^2 \times \mathbb{CP}^1$, which is $dP_1$.  Thus from the quiver we obtain homogeneous coordinates for $\mathbb{CP}^2\times \mathbb{CP}^1$ that describe
the $dP_1$ locus. The notation we have used makes the identification obvious.

Notice that in the quiver theory, the parameterizing variables $x_1,y_1,z,x_2,y_2$ are not homogeneous coordinates, so their rescalings are physical: they change the location of the branes. This is what gives us the cone structure for the total space. Notice also that the gauge invariants $\phi_{ij}$ are cubic in the parameterizing variables, so the moduli space of gauge invariants gives us
an embedding into $\mathbb{C}^9$. If we projectivize the embedding, we get the embedding into $\mathbb{CP}^8$ described above.
The relations between the monomials $\phi_{ij}$ are exactly the relations of this embedding\footnote{In general the $k$th del Pezzo surface, $dP_k$, which is $\mathbb{CP}^2$ blown up in $k$ points, admits an embedding $dP_k \hookrightarrow \mathbb{CP}^{9-k}$ for $k \leq 6$ (for $k >6$ we run out of variables since $\mathbb{CP}^2$ has only three variables). See \cite{GH} for more details.} since they are determined by the parameterizing variables.

To compute the volumes of interest, consider the divisor
\begin{equation}
H := f^{-1}\{ z =0 \},
\end{equation}
the exceptional divisor
\begin{equation}
 E := \left[ 0:0:1; x_2: y_2 \right] = \{ \text{a point} \} \times \mathbb{CP}^1,
 \end{equation}
and their difference
 \begin{equation}
 L:= H - E = \left[ 0:y_1:z_1;0:1 \right] \subset \mathbb{CP}^2 \times \{ \text{a point} \}.
 \end{equation}
The following table describes the zero loci of the various bifundamental fields; the results in the fourth column are derived below and we note that the last column follows from the fourth upon substituting $H =L+E$.
 
$$\begin{array}{|c||c|c|c|c|c|}
\hline
0 = & \begin{array}{c}\text{\scriptsize{so the only possible}}\\ \text{\scriptsize{nonzero coordinates are}} \end{array}& \text{\scriptsize{with constraints}} & \begin{array}{c}\text{\scriptsize{and thus the}}\\\text{\scriptsize{zero locus is}} \end{array} & \text{\scriptsize{or alternatively}}\\
\hline
\hline
\left\langle d \right\rangle & \left[ \phi_{20}: \phi_{21} \right] &\text{\scriptsize{none}} & E & E\\
\hline
z & \left[ \phi_{10}: \phi_{11}: \phi_{12}: \phi_{13} \right] & \phi_{10}\phi_{13} = \phi_{11}\phi_{12}, & H & L+E\\
&& \phi_{11}^2 = \phi_{12}\phi_{10}, &&\\
&& \phi_{12}^2 = \phi_{11}\phi_{13}.&&\\
\hline
x_2 & \left[ \phi_{02}:\phi_{13}: \phi_{21}\right] &  \phi_{02}^2 = \phi_{13}\phi_{21} & 2(H-E) & 2L\\
\hline
y_2 & \left[ \phi_{00}: \phi_{10}: \phi_{20} \right] & \phi_{00}^2 =\phi_{20}\phi_{10} & 2(H-E) & 2L\\
\hline
x_1 & \left[ \phi_{02}: \phi_{13}: \phi_{20}: \phi_{21} \right] & \phi_{02}^2 = \phi_{13}\phi_{21} & 2(H-E)+E & 2L+E\\
&&& = 2H -E & 
\\
\hline
y_1 & \left[ \phi_{00}: \phi_{10}: \phi_{20}: \phi_{21} \right]& \phi_{00}^2 = \phi_{10}\phi_{20} & 2(H-E)+E & 2L+E\\
&&&= 2H-E&\\
\hline
\end{array}$$
\indent By ``and thus the zero locus is'' we specifically mean inside $dP_1$ and not in the ambient space $\mathbb{CP}^8$. The multiplicities are computed by calculating the degrees of the appropriate locus in $\mathbb{CP}^8$. Notice that the condition that an arrow in the quiver vanishes implies that one of the variables above vanishes because of the gauge choices that are made. Thus the locus $a_2=0$ coincides with the locus $a_4=0$ for example.

Here we justify these computations.

Define the divisors $D_{x_i} := \{x_i=0\}$, $D_{y_i} := \{y_i = 0\}$, and $D_{z} := \{z=0\}$.  In order to write the classes of these divisors in terms of the basis $\{ \left[ L \right], \left[ E \right] \}$ of $H^1(dP_1)$, we compute their intersection numbers.  

 \begin{itemize}
  \item $D_{\left\langle d \right\rangle} = E$ since $\left\langle d \right\rangle = 0$ implies $x_1 = y_1 = 0$.
  \item $D_{z} \cdot E = 0$ since $z = 0$ implies $x_1 \not = 0$ or $y_1 \not = 0$, so $D_z$ does not intersect the exceptional divisor.  $D_z \cdot H = 1$ in $dP_1$ so $D_z =H$, though we note that in $\mathbb{CP}^8$ the zero locus $z=0$ is a twisted cubic and so $\deg(D_z) =3$. 
  \item  $x_2 =0$ implies $y_2 \not =0$ and since $y_1x_2 =x_1y_2$, it must be that $x_1 =0$.  But then the resulting coordinates $\left[ 0: y_1:x_1; 0:1 \right]$ are those of $L$.  Thus $D_{x_2} \cdot H = 2$ since $D_{x_2}$ is given by the degree 2 hypersurface $\phi_{02}^2 - \phi_{13}\phi_{21}=0$.  $D_{x_2} \cdot E = 2$ since again $D_{x_2}$ is a given by a degree 2 hypersurface and intersects $E$ transversely at the single point $\left[ 0:0:1;0:1\right]$.  Thus $D_{x_2} = 2(H -E)=2L$.  Similarly, $D_{y_2}=2(H-E)=2L$.
  \item The only difference between $D_{x_1}$ and $D_{x_2}$ is that $D_{x_1}$ is given by both coordinates $\left[ \phi_{20}:\phi_{21} \right]$ of the exceptional divisor whereas $D_{x_1}$ is only given by the single coordinate $\phi_{21}$.  Thus
  $$\{x_1=0\} = \{ x_2 =0\} \cup E,$$
and hence $D_{x_1} = D_{x_2}+E = 2L -E$.  Similarly, $D_{y_1}= 2L-E$.
\end{itemize}
 
We now consider the $R$-charges of the bifundamental fields.  These are determined by $a$-maximization \cite{IW} and were computed in \cite{BBC} (a general case was done in \cite{BHK2} for the toric phases of the $Y^{p,q}$ quivers for the case we need and we use their notation).
They are as follows:
\begin{equation} \begin{array}{rcl}
r(a_1) = r(b_1) & = & (3q-2p+\sqrt{4p^2-3q^2})/3q\\
 &=& \frac 13 \left( -1 + \sqrt{13} \right)\\
r(a_{2i}) =r(b_{2i})& = & 2p(2p-\sqrt{4p^2-3q^2})/3q^2\\
 & = & \frac 43 \left( 4 - \sqrt{13} \right)\\
r(c_i) & = & (-4p^2+3q^2+2pq+(2p-q)\sqrt{4p^2-3q^2})/3q^2\\
 & = & -3 + \sqrt{13}\\
r(d) &= &(-4p^2+3q^2-2pq+(2p+q)\sqrt{4p^2-3q^2})/3q^2\\
 & = & \frac 13 \left( -17+ 5 \sqrt{13} \right)
\end{array}
\end{equation}
Setting
\begin{equation}
r(c_i) :=  \int H  \ \ \text{ and } \ \ \ r(d) := \int E,
\end{equation}
we find
\begin{equation}\begin{array}{ccl}
r(a_{2i}) & = & \frac 43 \left( 4 - \sqrt{13} \right)\\
&= & \frac 13 \left( 6\left( -3+\sqrt{13} \right) + -2 \left( -17 + 5 \sqrt{13} \right)\right)\\
& =& 6 \left( \frac 13 r(c_i) \right) + -2 r(d) ,\\
&&\\
r(a_1) & = & \frac 13 \left( -1 + \sqrt{13} \right)\\
& = & \frac 13 \left( 6 \left( -3 + \sqrt{13} \right) -1 \left( -17 + 5 \sqrt{13} \right) \right)\\
& = & 6 \left( \frac 13 r(c_i) \right) - r(d),
\end{array}\end{equation}
and similarly for $b_{2i}$ and $b_1$.
These are exactly the same relations we get using the divisors described above. Moreover, as computed in \cite{BBC}, the volumes of the corresponding cycles in the Sasaki-Einstein manifold give the same dimensions. Here we see that the geometry we found matches exactly what is found in the AdS dual setup. Notice that our procedure was systematic and did not require any guesswork to make the match. Moreover all the multiplicities are accounted for by the algebraic geometry calculation.

It should also be noticed that a general analysis of D-branes in the $Y^{p,q}$  and $L^{a,b,c}$ geometries has been performed in \cite{Edelstein}, so it should be interesting to check the match between algebraic geometry and the geometry on the dual $AdS$ side for these examples. A recent analysis of how to get some handle on the geometry and some other studies of these operators can be found in \cite{EK, FGU}.

Also notice that although the examples we analyzed were given by toric geometries, in principle the techniques we have described do not depend on having so much symmetry. Instead, they only depend on the study of the details of the algebraic geometric space that one is considering. One should also notice that the property of a quiver field theory being toric is not preserved by Seiberg dualities \cite{DualC}, but the non-commutative algebraic geometry setup  is preserved \cite{BD}. This implies that any arrow in a quiver diagram can be interpreted as some element of $Hom(S_1, S_2)$ between two projective modules as described
by us. These are always sections of a global line bundle over the center of the algebras, which is common between them. One can think of the center variety as the moduli space of a point-like brane in the bulk, and remembering that the moduli spaces should be invariant under Seiberg dualities (see\cite{Brev} and also \cite{Asp} for more details on the geometry of points).

\section{A Seiberg dual of  the $\BC^3/\BZ_3$ orbifold}

The $\BC^3/\BZ_3$ quiver theory is usually represented by three nodes $e_1,e_2,e_3$, and between pairs of consecutive nodes one usually has three superfields $x_i,y_i,z_i$, giving a total of 9 arrows in the quiver.  Setting $x = \sum x_i$, $y= \sum y_i$, and $z = \sum z_i$ (the summation is formal, as described above), the F-terms can be deduced from the superpotential
\begin{equation}
W=\tr ([x,y]z)
\end{equation}
which is the same superpotential as ${\cal N}=4 $ SYM, properly projected to account for the three node quiver structure of the orbifold.

Since the variables $x,y,z$ commute with each other, any polynomial in these variables will also commute.  However, one can show that only cubic polynomials in the $x_i,y_i,z_i$ and their products can commute with the $e_i$.  The center is then generated by 
\begin{equation}
\alpha_{ijk} = x^i y^j z^k
\end{equation}
where $i+j+k=3$. The three variables $x,y,z$ have an associated $SU(3)$ symmetry of rotations between them. The cubic polynomials in $x,y,z$ transform as a ten dimensional totally symmetric representation of $SU(3)$. For irreducible representations, the vector space on each node is one dimensional, and so each arrow is represented by a scalar.  This is a special orbifold that has a toric description.

There is a one-to-one correspondence between the center of the algebra and the cycles at any given node. Thus again each node represents a line bundle on the Calabi-Yau cone. The cone is a complex cone over $\mathbb{CP}^2$.

The $\alpha_{ijk}$ give us a Veronese embedding of $\mathbb{CP}^2$ into $\mathbb{CP}^9$ very similar to the case of the first del Pezzo surface.
The basic dibaryon operators are objects like $\det( x_i)$. If $x_i$ vanishes, we find that
all the $\alpha_{ijk}=0$ if $i>0$, and the locus $x=0$ corresponds to a particular hyperplane section of the base of the cone. Also, the three non-gauge invariant objects $x_i,y_i,z_i$ at the node $i$ can
be understood as a set of homogeneous coordinates for $\mathbb{CP}^2$; they are global sections of the $\CO(1)$ line bundle on $\mathbb{CP}^2$. \typeout{The basic dibaryon operators are objects like $\det( x_i)$. If $x_i$ vanishes, we find that
all the $\alpha_{ijk}=0$ if $i>0$, and the locus $x_{i,i+1}=0$ corresponds to a particular hyperplane section of the base of the cone.}

The dibaryon operators for this orbifold theory have been analyzed by \cite{GRW}. The fact that there are three elementary dibaryons associated to the same hyperplane section on the $\mathbb{CP}^2$ base and that the lift to the $S^5/\BZ_3$ Sasaki-Einstein space is not simply connected suggested that each of these corresponds to a D-brane with different Wilson lines along the non-simply connected cycle.

Our idea is to show how one can redo the analysis of the basic dibaryons in a setting where the 
total space is not described as a toric variety from the field theory point of view. Our problem is to show how one can recover the geometric description of the baryonic operators above from a Seiberg dual theory \cite{Sdual}. According to the original arguments of Seiberg reproducing the counting of baryonic operators was given as evidence for the duality. Thus, we know it will work at this level. The question we will address is how to see the geometry from this more involved baryonic setup.

If we take the quiver diagram and do a Seiberg duality on any of the nodes, we get a new quiver diagram with three nodes.  We find that there are six arrows between two of the nodes, say $e_3$ and $e_2$. These should be thought of as transforming in the $6$ of $SU(3)$. We shall call them $\phi^{[ab]}$. The associated fields have dimension two. Moreover, there are also three arrows from $e_3$ to $e_1$ and three arrows from $e_1$ to $e_2$.  These transform in the $\bar 3$ representation of $SU(3)$ and have dimension $1/2$ (these computations can be found in \cite{DualC}). The reason they transform in opposite representations of $SU(3)$ is that the $SU(3)$ is part of the 'flavor symmetry' with respect to the gauge group node that was dualized. Remember that in a Seiberg duality one changes the direction of the arrows that begin and end on the node that is dualized (one trades fundamentals by antifundamentals plus mesons of the global symmetry). 

The superpotential is given by
\begin{equation}
W\sim \tr (\chi^{[\alpha\beta]} \phi_\alpha\xi_\beta 
\end{equation}
where the upper indices are fundamentals of $SU(3)$, and the lower indices are antifundamentals
(see \cite{Brev} for notation).

 For the irreducible representations of the algebra, we expect that the Seiberg duality corresponds to a derived equivalence between two different algebras 
 \cite{BD}, and that the points on the Calabi-Yau cone are skyscraper sheafs over the center of the original algebra that go to skyscarper sheafs over the center of the dualized algebra.

The arrows transforming in the $6$ of $SU(6)$ can be thought of as global sections of the $\CO(2)$ bundle on 
$\mathbb{CP}^2$, so one would expect a similar strategy as in the example over $Y^{2,1}$ and the conifold to deal with the corresponding dibaryons: one would obtain curves of degrees two on $\mathbb{CP}^2$ by studying general linear combinations of the bundles.

For the arrows transforming as a $\bar 3$ of $SU(3)$, these cannot be thought of as global sections for any line bundle on $\mathbb{CP}^2$. Line bundles on $\mathbb{CP}^2$ are described by the degree $d$. These generate a set of global sections that are in the $d$-symmetric product of $SU(3)$ (the d-symmetric product of global sections of $\CO(1)$).

Thus we find that we cannot interpret the arrows between nodes one and two, and between nodes two and three in terms of line bundles over $\mathbb{CP}^2$. This should not be so surprising. When one considers the irreducible representations of the algebra away from the singular locus, the node $e_1$ gets an associated  vector space of dimension two over the complex numbers. In physical setups one can argue this value from anomaly cancellation of the field theory, whereas at the other nodes one would have
spaces of dimension one. Thus, not all the vector spaces are the same dimension and one can have composition maps of arrows that give zero even though none of the arrows is zero itself. This would not be possible for sections of line bundles. What this means is that our prescription of choosing a zero locus of a global section of a line bundle needs to be generalized to a different construction that depends on having sections of general bundles rather than line bundles. After all, the spaces $Hom(\CAQ e_i, \CAQ e_j)$ will always be modules over the center and can always be interpreted in terms of coherent sheafs over the complement of the singular locus.

If one considers cycles based at node $e_2$, one finds that all the minimal cycles that can be non-vanishing belong to a ten dimensional representation of $SU(3)$ \cite{Brev}, and that they all commute with each other in a trivial way on all irreducible representations of the algebra (since they are all $1\times 1$ matrices, this is, complex numbers). As a vector space one realizes that this is the same as the vector space of the $\alpha_{ijk}$ that we described before. Thus one sees indirectly that the center of the algebra is invariant under the derived equivalence.

So how do we build invariants if we have volume forms attached to each node, and they have different dimension? Remember that in the conformal field theory the rank of the gauge group at $e_1$ will be double the rank of the gauge group at $e_2$. Let $V_1 = \mathbb{C}^{2N}$ be the vector space on which the gauge group at $e_1$ acts, and similarly let $V_2 = \mathbb{C}^{N}$ be the vector space on which the gauge group at $e_2$ acts.  We first consider linear maps $V_2 \rightarrow V_1$, that is, representations of the arrows from node $e_2$ to node $e_1$.  Since the dimension of $V_2$ will be half that of $V_1$, by considering two such maps $\gamma,\delta: V_2 \rightarrow V_1$, we can generate in the general case two subvector spaces at node $e_1$ of dimension $N$. If we pushforward the volume form of $e_2$ via each of these maps, we get two elements of the vector space $\omega_{\gamma}, \omega_{\delta} \in \Lambda^N V_1$. Via the wedge product of these two, we can find a unique number associated to these two maps, namely $\omega_{\gamma}\wedge \omega_{\delta}\in \Lambda^{2N} V_1$ which can now be compared with the volume form at node $e_1$.
Thus we can produce a single complex number given this information. This is the evaluation map of a dibaryon operator on the given configuration.

On the moduli space, the maps split into irreducible representations, so these objects factorize between irreducibles, and for each irreducible we find a similar object with $N=1$. If we use two maps in the 
$\bar 3$ representation, because of the wedge product, we find an antisymmetric object in the two entries with specific transformations under the $SU(3)$ symmetry. These transform as a $3$ of $SU(3)$ and can again be interpreted as a section of a line bundle (namely $\CO(1)$ on $\mathbb{CP}^2$). The locus associated to the vanishing of this object is when the pair of maps $\gamma, \delta$
are degenerate (they give the same one dimensional subspace). This can also be written as follows.

Consider the map from $V_2\oplus V_2\to V_1$ given by
$\mu(v, u) = \gamma(v)+\delta(u)$. The object we have described above is 
what one would ordinarily write as $\det(\mu)$. This vanishes if the map is not invertible, and is 
exactly when the map $\mu$ has a jump in the dimension of the kernel (or one could phrase it in terms of a jump of the dimension of the co-kernel as well). Notice also that this is a larger locus than the vanishing of any individual arrow.

The corresponding determinant can be written as the following gauge invariant combination of fields
\begin{equation}
\det(\mu) = \epsilon_{i_1, \dots i_N} \epsilon_{j_1,\dots, j_N} X^{i_1}_{k_1}\dots X^{i_N}_{k_N}
 Y^{j_1}_{k_{N+1}} \dots Y^{j_N}_{k_{2N}} \epsilon^{k_1\dots k_{2N}}
\end{equation}
and as we argued, this factorizes on the moduli space to a product of sections of line bundles for each individual irreducible. However, the arrows themselves have to be interpreted here as global sections of
general bundles that can degenerate. In this case, one can argue that the pair $(X,Y)$ belongs to $Hom(V_2\oplus V_2 , V_1)$ and that since we have spaces of dimension two on the irreducibles, the $(X,Y)$ are naturally global sections of  a sheaf of matrix-valued $2\times 2$ matrices over the center. The volume of the singular locus determined by these maps gives a holomorphic submanifold of the base of the cone that we identify with the $R$-charge of the multibaryon operator.

Notice that in the Seiberg dual the dimension of what we called $X,Y$ are one half, and that the object we have described has the same dimension as the dibaryons of the original orbifold quiver theory. Moreover, we have argued that the same locus on the geometry of the base is associated to the two theories.

One can similarly construct the other baryons of the right dimension. If one considers composite maps from $V_2 \to V_3$, the F-terms also tell us that on an irreducible these maps transform as a $3$ of $SU(3)$. Similarly, one can consider general maps $V_1\to V_3\oplus V_3$ to define the third class of elementary baryon of dimension $N= 2N * 1/2$.

{\bf Acknowledgements:} We would like to thank S. Franco and D. Morrison for many discussions related to this work. Work supported in part by DOE under grant DE-FG02-91ER40618. D. B. would like to thank the hospitality of the Kavli Institute of Theoretical Physics China for their support.


\begin{thebibliography}{99}
\bibitem{Malda}
  J.~M.~Maldacena,
  ``The large N limit of superconformal field theories and supergravity,''
  Adv.\ Theor.\ Math.\ Phys.\  {\bf 2}, 231 (1998)
  [Int.\ J.\ Theor.\ Phys.\  {\bf 38}, 1113 (1999)]
  [arXiv:hep-th/9711200].

\bibitem{FHMSVW}
  S.~Franco, A.~Hanany, D.~Martelli, J.~Sparks, D.~Vegh and B.~Wecht,
  ``Gauge theories from toric geometry and brane tilings,''
  JHEP {\bf 0601}, 128 (2006)
  [arXiv:hep-th/0505211].


\bibitem{Vafa}
  F.~Cachazo, S.~Katz and C.~Vafa,
  ``Geometric transitions and N = 1 quiver theories,''
  arXiv:hep-th/0108120.
  F.~Cachazo, B.~Fiol, K.~A.~Intriligator, S.~Katz and C.~Vafa,
  ``A geometric unification of dualities,''
  Nucl.\ Phys.\  B {\bf 628}, 3 (2002)
  [arXiv:hep-th/0110028].
  

\bibitem{VW}
  C.~Vafa,
  ``Modular Invariance And Discrete Torsion On Orbifolds,''
  Nucl.\ Phys.\  B {\bf 273}, 592 (1986).
  C.~Vafa and E.~Witten,
  ``On orbifolds with discrete torsion,''
  J.\ Geom.\ Phys.\  {\bf 15}, 189 (1995)
  [arXiv:hep-th/9409188].
  
\bibitem{Brev}
  D.~Berenstein,
  ``Reverse geometric engineering of singularities,''
  JHEP {\bf 0204}, 052 (2002)
  [arXiv:hep-th/0201093].
 
\bibitem{BJL}
  D.~Berenstein, V.~Jejjala and R.~G.~Leigh,
  ``Marginal and relevant deformations of N = 4 field theories and
  non-commutative moduli spaces of vacua,''
  Nucl.\ Phys.\  B {\bf 589}, 196 (2000)
  [arXiv:hep-th/0005087].
  D.~Berenstein and R.~G.~Leigh,
  ``Resolution of stringy singularities by non-commutative algebras,''
  JHEP {\bf 0106}, 030 (2001)
  [arXiv:hep-th/0105229].

\bibitem{Seiberg}
  N.~Seiberg,
  ``Exact Results On The Space Of Vacua Of Four-Dimensional Susy Gauge
  Theories,''
  Phys.\ Rev.\  D {\bf 49}, 6857 (1994)
  [arXiv:hep-th/9402044].
  N.~Seiberg,
  ``The Power of holomorphy: Exact results in 4-D SUSY field theories,''
  arXiv:hep-th/9408013.

\bibitem{WGKP}
  S.~S.~Gubser, I.~R.~Klebanov and A.~M.~Polyakov,
  ``Gauge theory correlators from non-critical string theory,''
  Phys.\ Lett.\  B {\bf 428}, 105 (1998)
  [arXiv:hep-th/9802109].
  E.~Witten,
  ``Anti-de Sitter space and holography,''
  Adv.\ Theor.\ Math.\ Phys.\  {\bf 2}, 253 (1998)
  [arXiv:hep-th/9802150].
  
\bibitem{FR}
  P.~G.~O.~Freund and M.~A.~Rubin,
  ``Dynamics Of Dimensional Reduction,''
  Phys.\ Lett.\  B {\bf 97}, 233 (1980).
  
\bibitem{MP}
  D.~R.~Morrison and M.~R.~Plesser,
  ``Non-spherical horizons. I,''
  Adv.\ Theor.\ Math.\ Phys.\  {\bf 3}, 1 (1999)
  [arXiv:hep-th/9810201].

\bibitem{Wittenbar}
  E.~Witten,
  ``Baryons and branes in anti de Sitter space,''
  JHEP {\bf 9807}, 006 (1998)
  [arXiv:hep-th/9805112].
  
\bibitem{GK}
  S.~S.~Gubser and I.~R.~Klebanov,
  ``Baryons and domain walls in an N = 1 superconformal gauge theory,''
  Phys.\ Rev.\  D {\bf 58}, 125025 (1998)
  [arXiv:hep-th/9808075].
  
\bibitem{Mikhailov}
  A.~Mikhailov,
  ``Giant gravitons from holomorphic surfaces,''
  JHEP {\bf 0011}, 027 (2000)
  [arXiv:hep-th/0010206].


\bibitem{Beasley}
  C.~E.~Beasley,
  ``BPS branes from baryons,''
  JHEP {\bf 0211}, 015 (2002)
  [arXiv:hep-th/0207125].
  
\bibitem{IW}
  K.~A.~Intriligator and B.~Wecht,
  ``Baryon charges in 4D superconformal field theories and their AdS duals,''
  Commun.\ Math.\ Phys.\  {\bf 245}, 407 (2004)
  [arXiv:hep-th/0305046].
  C.~P.~Herzog and J.~McKernan,
  ``Dibaryon spectroscopy,''
  JHEP {\bf 0308}, 054 (2003)
  [arXiv:hep-th/0305048].

\bibitem{BBC}
  M.~Bertolini, F.~Bigazzi and A.~L.~Cotrone,
  ``New checks and subtleties for AdS/CFT and a-maximization,''
  JHEP {\bf 0412}, 024 (2004)
  [arXiv:hep-th/0411249].
  
\bibitem{IW1}
  K.~A.~Intriligator and B.~Wecht,
  ``The exact superconformal R-symmetry maximizes a,''
  Nucl.\ Phys.\  B {\bf 667}, 183 (2003)
  [arXiv:hep-th/0304128].

\bibitem{BFZ}
  A.~Butti, D.~Forcella and A.~Zaffaroni,
 ``Counting BPS baryonic operators in CFTs with Sasaki-Einstein duals,''
  JHEP {\bf 0706}, 069 (2007)
  [arXiv:hep-th/0611229].

\bibitem{KW}
  I.~R.~Klebanov and E.~Witten,
``Superconformal field theory on threebranes at a Calabi-Yau  singularity,''
  Nucl.\ Phys.\  B {\bf 536}, 199 (1998)
  [arXiv:hep-th/9807080].

\bibitem{BHK}
  D.~Berenstein, C.~P.~Herzog and I.~R.~Klebanov,
  ``Baryon spectra and AdS/CFT correspondence,''
  JHEP {\bf 0206}, 047 (2002)
  [arXiv:hep-th/0202150].

\bibitem{WBagger}
  J.~Wess and J.~Bagger,
  ``Supersymmetry and supergravity,''
{\it  Princeton, USA: Univ. Pr. (1992) 259 p}

\bibitem{Bcon}
  D.~Berenstein,
  ``On the universality class of the conifold,''
  JHEP {\bf 0111}, 060 (2001)
  [arXiv:hep-th/0110184].

\bibitem{FHH}
  S.~Benvenuti, B.~Feng, A.~Hanany and Y.~H.~He,
  ``Counting BPS operators in gauge theories: Quivers, syzygies and
  plethystics,''
  JHEP {\bf 0711}, 050 (2007)
  [arXiv:hep-th/0608050].
  D.~Martelli and J.~Sparks,
  ``Dual giant gravitons in Sasaki-Einstein backgrounds,''
  Nucl.\ Phys.\  B {\bf 759}, 292 (2006)
  [arXiv:hep-th/0608060].
  B.~Feng, A.~Hanany and Y.~H.~He,
  ``Counting gauge invariants: The plethystic program,''
  JHEP {\bf 0703}, 090 (2007)
  [arXiv:hep-th/0701063].


\bibitem{LT}
  M.~A.~Luty and W.~Taylor,
  ``Varieties of vacua in classical supersymmetric gauge theories,''
  Phys.\ Rev.\  D {\bf 53}, 3399 (1996)
  [arXiv:hep-th/9506098].


\bibitem{Master}
  D.~Forcella, A.~Hanany, Y.~H.~He and A.~Zaffaroni,
  ``The Master Space of N=1 Gauge Theories,''
  JHEP {\bf 0808}, 012 (2008)
  [arXiv:0801.1585 [hep-th]].


\bibitem{Douglas}
  M.~R.~Douglas,
  ``D-branes, categories and N = 1 supersymmetry,''
  J.\ Math.\ Phys.\  {\bf 42}, 2818 (2001)
  [arXiv:hep-th/0011017].



\bibitem{BHart}
  D.~Berenstein,
  ``Large N BPS states and emergent quantum gravity,''
  JHEP {\bf 0601}, 125 (2006)
  [arXiv:hep-th/0507203].
  D.~Berenstein,
  ``Strings on conifolds from strong coupling dynamics, part I,''
  JHEP {\bf 0804}, 002 (2008)
  [arXiv:0710.2086 [hep-th]].
  D.~E.~Berenstein and S.~A.~Hartnoll,
  ``Strings on conifolds from strong coupling dynamics: quantitative results,''
  JHEP {\bf 0803}, 072 (2008)
  [arXiv:0711.3026 [hep-th]].

\bibitem{GMSW}
  J.~P.~Gauntlett, D.~Martelli, J.~Sparks and D.~Waldram,
  ``Sasaki-Einstein metrics on S(2) x S(3),''
  Adv.\ Theor.\ Math.\ Phys.\  {\bf 8}, 711 (2004)
  [arXiv:hep-th/0403002].
  D.~Martelli and J.~Sparks,
  ``Toric geometry, Sasaki-Einstein manifolds and a new infinite class of
  AdS/CFT duals,''
  Commun.\ Math.\ Phys.\  {\bf 262}, 51 (2006)
  [arXiv:hep-th/0411238].





\bibitem{FHHtoric}
  B.~Feng, A.~Hanany and Y.~H.~He,
  ``D-brane gauge theories from toric singularities and toric duality,''
  Nucl.\ Phys.\  B {\bf 595}, 165 (2001)
  [arXiv:hep-th/0003085].

\bibitem{HW}
  C.~P.~Herzog and J.~Walcher,
  ``Dibaryons from exceptional collections,''
  JHEP {\bf 0309}, 060 (2003)
  [arXiv:hep-th/0306298].

\bibitem{BHOP}
  D.~Berenstein, C.~P.~Herzog, P.~Ouyang and S.~Pinansky,
  ``Supersymmetry breaking from a Calabi-Yau singularity,''
  JHEP {\bf 0509}, 084 (2005)
  [arXiv:hep-th/0505029].


\bibitem{AFHS}
  B.~S.~Acharya, J.~M.~Figueroa-O'Farrill, C.~M.~Hull and B.~J.~Spence,
  ``Branes at conical singularities and holography,''
  Adv.\ Theor.\ Math.\ Phys.\  {\bf 2}, 1249 (1999)
  [arXiv:hep-th/9808014].


\bibitem{B1} C. Beil, ``The noncommutative geometry of the $Y^{p,q}$ quivers,'' in preparation.

\bibitem{GH} P. Griffiths and J. Harris, ``Principles of algebraic geometry,'' Wiley, 1978.

\bibitem{BHK2}
  S.~Benvenuti, A.~Hanany and P.~Kazakopoulos,
  ``The toric phases of the Y(p,q) quivers,''
  JHEP {\bf 0507}, 021 (2005)
  [arXiv:hep-th/0412279].

\bibitem{Edelstein}
  F.~Canoura, J.~D.~Edelstein, L.~A.~P.~Zayas, A.~V.~Ramallo and D.~Vaman,
  ``Supersymmetric branes on AdS(5) x Y**(p,q) and their field theory  duals,''
  JHEP {\bf 0603}, 101 (2006)
  [arXiv:hep-th/0512087].
  F.~Canoura, J.~D.~Edelstein and A.~V.~Ramallo,
  ``D-brane probes on L(a,b,c) superconformal field theories,''
  JHEP {\bf 0609}, 038 (2006)
  [arXiv:hep-th/0605260].

\bibitem{EK}
  J.~Evslin and S.~Kuperstein,
  ``Trivializing a Family of Sasaki-Einstein Spaces,''
  arXiv:0803.3241 [hep-th].
  J.~Evslin and S.~Kuperstein,
  ``Which BPS Baryons Minimize Volume?,''
  arXiv:0809.4594 [hep-th].


\bibitem{FGU}
  D.~Forcella, I.~Garcia-Etxebarria and A.~Uranga,
  ``E3-brane instantons and baryonic operators for D3-branes on toric
  singularities,''
  arXiv:0806.2291 [hep-th].
  
\bibitem{DualC}
  A.~Hanany and J.~Walcher,
  ``On duality walls in string theory,''
  JHEP {\bf 0306}, 055 (2003)
  [arXiv:hep-th/0301231].

\bibitem{BD}
  D.~Berenstein and M.~R.~Douglas,
  ``Seiberg duality for quiver gauge theories,''
  arXiv:hep-th/0207027.
  
  
\bibitem{Asp}
  P.~S.~Aspinwall,
  ``A point's point of view of stringy geometry,''
  JHEP {\bf 0301}, 002 (2003)
  [arXiv:hep-th/0203111].
  
  
\bibitem{GRW}
  S.~Gukov, M.~Rangamani and E.~Witten,
  ``Dibaryons, strings, and branes in AdS orbifold models,''
  JHEP {\bf 9812}, 025 (1998)
  [arXiv:hep-th/9811048].
  
\bibitem{Sdual}
  N.~Seiberg,
  ``Electric - magnetic duality in supersymmetric nonAbelian gauge theories,''
  Nucl.\ Phys.\  B {\bf 435}, 129 (1995)
  [arXiv:hep-th/9411149].
 
 
 
  
\end{thebibliography}
\end{document}